\documentclass{jps-cp}
\usepackage{txfonts} %Please comment out this line unless the txfonts package is availabe in your LaTeX system.

\title{Electronic phase diagram in Te-annealed superconducting FeTe$_{1-x}$Se$_x$ revealed by magnetic susceptibility}

\author{Takenori \textsc{Fujii}$^{1}$, Yu \textsc{Uezono}$^{2}$, Takumi \textsc{Otsuka}$^{2}$, Shotaro \textsc{Hagisawa}$^{2}$ and Takao \textsc{Watanabe}$^{2}$}

\inst{$^{1}$Cryogenic Research Center, The University of Tokyo, Bunkyo, Tokyo 113-0032, Japan\\
$^{2}$Graduate School of Science and Technology, Hirosaki University, Hirosaki 36-8561, Japan}

\email{fujii@crc.u-tokyo.ac.jp}

\recdate{July 15, 2022}

\abst{Knowledge of the doping ($x$)-temperature ($T$) phase diagram of Fe-based superconductors is crucial in understanding the mechanism of high transition temperature superconductivity. Here, we measured the Se doping dependence of the magnetic susceptibility for Te-annealed FeTe$_{1-x}$Se$_x$. Two characteristic temperatures $T^{*}_{\chi}$ and $T^{**}_{\chi}$ were observed in the temperature dependence of magnetic susceptibility, where the magnetic susceptibility decreased below $T^{**}_{\chi}$ and tended to increase with further decreasing temperature below $T^{*}_{\chi}$. The decrease in magnetic susceptibility below $T^{**}_{\chi}$ is considered to be due to the opening of the pseudogap. Whereas the increase in magnetic susceptibility below $T^{*}_{\chi}$ is attributed to the increase in the carrier concentration of electrons. Furthermore, we found the large superconducting fluctuation, which may be related to BCS-BEC crossover.}

\kword{FeTe$_{1-x}$Se$_x$ single crystals, Te-annealing, normal state magnetic susceptibility, pseudogap, crossover from the incoherent to the coherent electronic state, BCS-BEC crossover}

\begin{document}
\maketitle

\section{Introduction}
Normal state properties of iron-based superconductors have been crucial in understanding the mechanism of superconductivity, and they have been garnering increasing attention, since the discovery of their high superconducting transition temperature $T_c$ \cite{firstreport}. Various problems with this material, such as the anomalous non-Fermi-liquid-like normal-state transport properties\cite{Fermilq} and the possibility of the Bardeen-Cooper-Shrieffer (BCS) - Bose-Einstein condensation (BEC) crossover regime\cite{BCS-BEC01,BCS-BEC02,BCS-BEC03,BCS-BEC04,BCS-BEC05}, are particularly in the forefront of condensed matter physics. Among the several Fe-based superconductors, the FeTe$_{1-x}$Se$_x$ system has an advantage in addressing the aforementioned problems because its crystal structure is the simplest, i.e, it comprises only conducting FeX (X: Te or Se) layers. However, excess Fe is inevitably incorporated in the crystals, which conceals the nature of anomalous physical properties in FeTe$_{1-x}$Se$_x$ systems. Hence, a lot of efforts have been made to eliminate obstructive excess Fe\cite{anneal01,anneal02,anneal03,anneal04}. In our previous research, we developed an annealing method that discards excess Fe by annealing single crystals in tellurium vapor (Te anneal)\cite{anneal05}. Using this Te-annealing method, the excess Fe was sufficiently eliminated without damaging the samples. 
With this high quality sample, we performed the magneto-transport measurements and inferred that the crossover from the incoherent to the coherent electronic state and the opening of the pseudogap occur at high temperatures\cite{MR}.  Furthermore, via magnetoresistance measurements, we determined that the temperature $T_{scf}$ at which the superconducting fluctuations occur was 2.7 times larger than $T_c$, which is consistent with the behavior of the BCS-BEC crossover regime\cite{LTp}. 

The evolution from incoherent to coherent electronic states with increasing Se doping was also observed by angle-resolved photoemission spectroscopy (ARPES), which elucidates the close relationship between the coherent electronic state and emergence of superconductivity\cite{coherent01,coherent02}. On the other hand, whether iron-based superconductors, especially FeSe and FeSe$_{1-x}$S$_x$ are in the BCS-BEC crossover regime or not is hotly debated relative to the giant superconducting fluctuations\cite{BCS-BEC01,BCS-BEC02,BCS-BEC03,BCS-BEC04,BCS-BEC05}. 
Therefore, clarifying the doping ($x$)-temperature ($T$) phase diagram of Fe-based superconductors by using various measurements is important in elucidating the mechanism of superconductivity. 
Here, we measured the normal state magnetic susceptibility of Te annealed FeTe$_{1-x}$Se$_x$ ($x$ = 0.2, 0.3, and 0.4) and investigated the doping ($x$)-temperature ($T$) phase diagram.

\section{Experiment}
Single crystals of FeTe$_{1-x}$Se$_x$ were grown using the Bridgman method\cite{bridgeman} with their nominal compositions. As-grown crystals were cleaved into thin crystals with 1-mm thickness, and they were sealed into pyrex or quartz tube with pulverized Te. Then, the tube was heated for more than 400 h at 400 $^\circ$C. Annealed crystals were usually covered with grayish accretion. However, by carefully removing the accretion, shiny and flat surface emerged. The superconducting transition of the annealed crystals were very sharp ($\Delta T_c \leq $ 1K), thus indicating that the excess iron was completely discarded. Magnetic susceptibility was measured using a superconducting quantum interference device (SQUID) magnetometer (Quantum Design MPMS3) with the magnetic field applied parallel to the thin sample.

\section{Results and Discussion}
\subsection{Impurity phase of Fe$_3$O$_4$ in the Te-annealed samples}
Figure 1(a) presents the temperature dependence of the magnetic susceptibility for as-grown FeTe$_{0.6}$Se$_{0.4}$. It exhibits Curie-Weiss-like behavior at low temperatures and slight symptoms of superconductivity below 5K. This Curie-Weiss-like behavior is attributed to the local moment from excess Fe; in addition, superconductivity is destroyed by the local moment. Below 50 K, the difference in susceptibility at each magnetic field is evident. The inset of Figure 1(a) illustrates the magnetic field dependence of the magnetic moment ($M-H$ plot).  It presents the linear field dependence at high temperatures, while the deviation from the linear field dependence was observed at 10 K. These results indicate that the excess Fe works as local moment at high temperatures and that the ferromagnetic interaction of excess Fe would develop below 50 K. Figure 1(b) presents the temperature dependence of the magnetization for Te-annealed  FeTe$_{0.6}$Se$_{0.4}$ measured at 7 T. It exhibits an almost temperature independent behavior, which implies that Pauli paramagnetism is dominant. The inset of Figure 1(b) presents a $M-H$ plot for Te-annealed  FeTe$_{0.6}$Se$_{0.4}$. We can clearly observe the ferromagnetism, which could have emerged from the impurity phase other than excess Fe. To extract the Ferromagnetic component from the obtained susceptibility of Te-annealed sample, we deduced residual magnetic moment at 0 Oe, which is the $y$-intercept of linear magnetic moment in $M-H$ plot. 
Using the temperature dependence of the magnetic moment at 7 T ($M_{7T}(T)$) and 1 T ($M_{1T}(T)$), the residual magnetization $M_{res}(T)$ was calculated using the equation,

\begin{equation}
M_{res}(T) = M_{1T}(T) - \frac{M_{7T}(T) - M_{1T}(T)}{6} \times 1. 
\end{equation}
Figure 1(c) presents the calculated residual magnetic moment, which emerges from the impurity phase. As observed in the figure, clear phase transition can be detected at approximately 120 K. This is considered a Verwey transition in magnetite (Fe$_3$O$_4$)\cite{Verwey}. Note that, there are no symptoms of transition around 120 K in the original FeTe$_{0.6}$Se$_{0.4}$ data, which suggests that the inclusion of the impurity phase (Fe$_3$O$_4$) was negligible. Because the saturated magnetization of magnetite is known to be approximately 90 emu/g at room temperature, we can estimate the impurity phase amount of magnetite as 2.46 $\times$ 10$^{-3}$ mg in 5.34 mg FeTe$_{0.6}$Se$_{0.4}$ (0.046 \%). The amount of the impurity phase does not depend on Se doping, but on the position of the samples. Because this impurity phase is absent in the as-grown samples, it is considered to have been introduced during the annealing process. In the following analysis, this ferromagnetic component was subtracted from the raw data. However, our conclusion was not altered depending on the details of subtraction.

\subsection{Estimation of characteristic temperatures}
Figure 2(a) illustrates the temperature dependence of magnetic susceptibility of the samples with various Se doping. It exhibits a rather large susceptibility (4 - 9 $\times$ 10$^{-6}$ emu/gOe), which is 10 times larger than that of high-$Tc$ cuprate Bi$_2$Sr$_2$CaCu$_2$O$_{8-\delta}$\cite{watanabe}. The observed large susceptibility is consistent with a previous report\cite{sus}. The absolute values of the magnetic susceptibility increase drastically with decreasing Se contents. Because Pauli paramagnetism would be dominant in the magnetic susceptibility in FeTe$_{1-x}$Se$_x$, the increase in the magnetic susceptibility implies the increase in DOS at the Fermi level. In fact, such an increase in DOS was observed in the specific heat measurement, where, the electronic specific-heat coefficient in the normal state $\gamma_n$ increases as Se decreases\cite{HC}. Because the carrier density would not change drastically (Te is an isovalent substitution), the increase in the effective mass due to strongly correlated electron would be realized toward a quantum critical point between the antiferromagnetic and superconducting phases ($\sim$ $x$ = 0.07). 

The overall temperature dependence of the magnetic susceptibility in FeTe$_{1-x}$Se$_x$ is as follows. With decreasing temperature, the magnetic susceptibility decreases below $T^{**}_{\chi}$ (indicated by red arrows). With further decreasing temperature, slight upturn can be observed below $T^{*}_{\chi}$ (indicated by blue arrows). To estimate the characteristic temperatures $T^{*}_{\chi}$, $T^{**}_{\chi}$, and $T_{scf}$ in the magnetic susceptibility, we plotted the temperature derivative of magnetic susceptibility vs temperature for $x$ = 0.2, 0.3, and 0.4 in Figure 2(b), 2(c), and 2(d) respectively. As observed in these figures, the temperature derivative changes its slope twice with decreasing temperature. Hence, we present the solid straight lines that are linear extrapolations at each temperature ranges. $T^{*}_{\chi}$ and $T^{**}_{\chi}$ were defined from the intersection point of the extrapolation lines.  Similarly, $T_{scf}$ was determined by the temperature from which the temperature derivative start to deviate from the linear extrapolation line. The obtained temperatures  are consistent with the observations of our magnetotransport measurements\cite{MR}. 

Figure 3 presents the $x-T$ phase diagram for FeTe$_{1-x}$Se$_x$, including the data from a previous study\cite{MR}. 
$T^{**}_{\chi}$ corresponds to $T^*_{\rho ab}$ at which $\rho_{ab}$ reaches its broad maximum. The decrease in magnetic susceptibility below $T^{**}_{\chi}$ is considered to be due to the decrease in DOS by opening the pseudogap. In strongly correlated iron chalcogenides, such as FeTe$_{1-x}$Se$_x$, the existence of an orbital-selective Mott phase (OSMP) and related incoherent to coherent transition were theoretically proposed\cite{OSMP01,OSMP02}. As explained in a previous report\cite{MR}, $T^{**}_{\chi}$ would correspond to the transformation from OSMP to the metallic state. Recently, a similar phase diagram was proposed using the ARPES measurement\cite{PhaseDiagram}. When the states become coherent, the Fermi surfaces become well defined with some type of band hybridization, which would cause the pseudogap to open via interband nesting\cite{MR}. On the other hand, another characteristic temperature $T^{*}_{\chi}$ appears to correspond with $T^*_{\rho c}$, below which $\rho_{c}$ values exhibit the typical plateau. As mentioned in a previous report, the electron carriers would participate in charge transport below this temperature; hence, magnetic susceptibility would increase due to the increase in DOS\cite{MR}. As observed in Fig. 2(a), the magnetic susceptibility decreased again below $T_{scf}$ when the temperature was further decreased (indicated by green arrows). We consider that the decrease in magnetic susceptibility is attributed to the diamagnetism originated in superconducting fluctuation. The estimated $T_{scf}$ values from Fig. 2(b)-(d) were $\sim$ 57 K for $x$ = 0.4, $\sim$ 45 K for $x$ = 0.3, and $\sim$ 31 K for $x$ = 0.2. Notably, they were 3.9, 3.2, and 2.4 times larger than their corresponding $T_c$ for $x$ = 0.4, $x$ = 0.3, and $x$ = 0.2 respectively.  In the BCS-BEC crossover regime, the effect of superconducting fluctuations is expected to be enhanced, as Cooper pairs can be formed at higher temperatures than $T_c$\cite{BCS-BEC01}. These results indicate that the FeTe$_{1-x}$Se$_x$ system is indeed in the BCS-BEC crossover regime.

\section{Conclusion}
To estimate characteristic temperatures, such as $T^{**}_{\chi}$, $T^{*}_{\chi}$, and $T_{scf}$, we measured the magnetic susceptibility for high-quality Te-annealed  FeTe$_{1-x}$Se$_x$. The absolute value of magnetic susceptibility increased as Se contents decreased, which is considered to be due to the increase in DOS as it approaches a quantum critical point between the antiferromagnetic and superconducting phases. We observed the decrease in magnetic susceptibility below $T^{**}_{\chi}$, which is considered to be due to the opening of the pseudogap. On the other hand, the increase in magnetic susceptibility below $T^{*}_{\chi}$ is attributed to the increase in DOS due to the participant of electron carrier. These results agree with our previous magnetotransport study\cite{MR}. Moreover, we inferred that $T_{scf}$ values were substantially high and they reached 57, 45, and 31 K for $x$ = 0.4, 0.3, and 0.2 respectively. In particular, it was 3.9 times larger than  its $T_c$ for $x$ = 0.4. These high $T_{scf}$ values are consistent with the behavior of the BCS-BEC crossover regime.

In this study, we determined that our Te-annealed sample contained a ferromagnetic impurity phase. Although the amount of this impurity phase is negligible, it may conceal the intrinsic behaviors of superconductivity at a low field. In the future, we have to improve the annealing process, to completely eliminate ferromagnetic impurity and reveal more detailed magnetic properties.

\section*{Acknowledgment}
This work was supported by JSPS KAKENHI Grant No. 20K03849 and the Hirosaki University Grant for Distinguished Researchers from the fiscal year 2017 to 2018. 
\newpage

\section*{Figure captions}
\begin{figure}[tbh]
\begin{center}
\end{center}
\caption{(a): Temperature dependence of the magnetic susceptibility for as-grown sample of $x$ = 0.4. (b): Temperature dependence of the magnetization for Te-annealed sample of $x$ = 0.4 measured at 7 T. The blue line presents the data that the ferromagnetic component from Fe$_3$O$_4$ was subtracted. Inset of Fig 1(a) and (b):  Magnetic field dependence of the magnetization. (c): Temperature dependence of calculated residual magnetic moment that indicates the magnetic moment of the impurity phase magnetite (Fe$_3$O$_4$).}
\label{f1}
\end{figure}
\begin{figure}[tbh]
\begin{center}
\end{center}
\caption{(a): Temperature dependence of the magnetic susceptibility of the Te-annealed sample with various Se doping measured at 7 T. (b), (c), (d): Temperature derivative of the magnetic susceptibility vs temperature for $x$ = 0.2, 0.3, and 0.4 respectively. The solid straight lines represent linear extrapolations at a certain temperature range, which are presented as a guideline.}
\label{f2}
\end{figure}
\begin{figure}[tbh]
\begin{center}
\end{center}
\caption{Characteristic temperatures $T^{*}_{\chi}$, $T^{**}_{\chi}$, and $T_{scf}$ vs Se concentration $x$ for Te-annealed  FeTe$_{1-x}$Se$_x$, plotted together with data from previous magneto transport study\cite{MR}. The open circles of $T^{*}_{RH}$ are replotted from Ref. \cite{sun}.}
\label{f3S}
\end{figure}

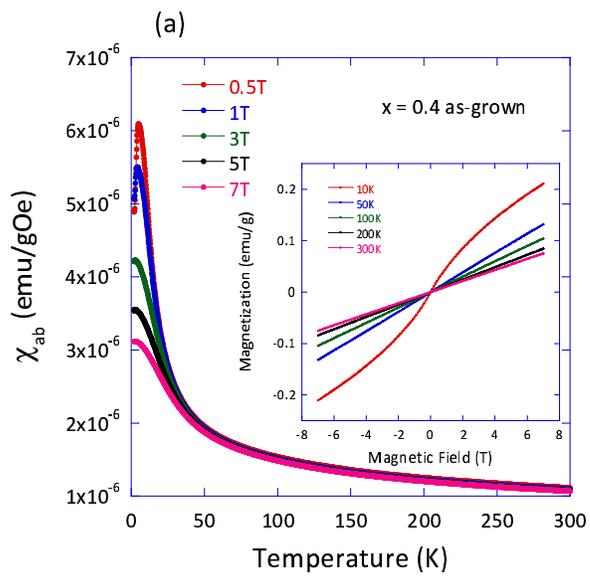
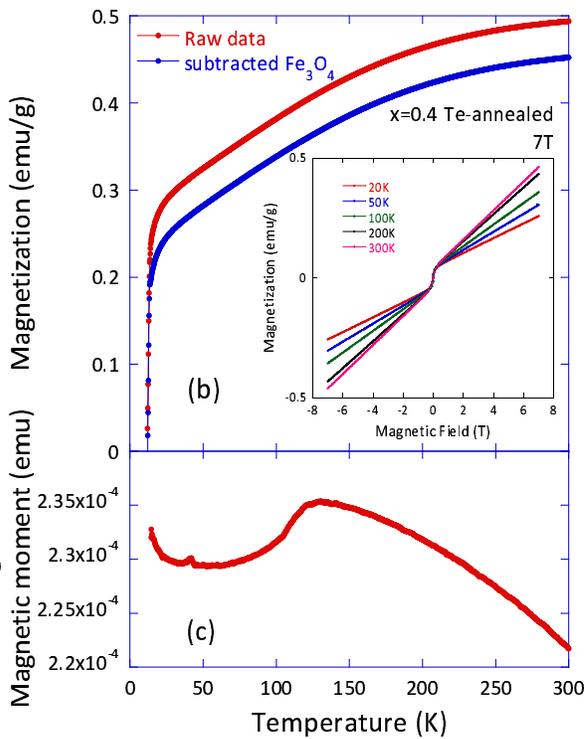
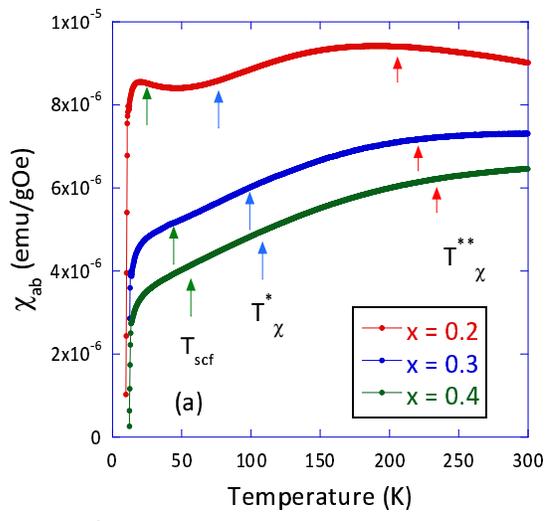
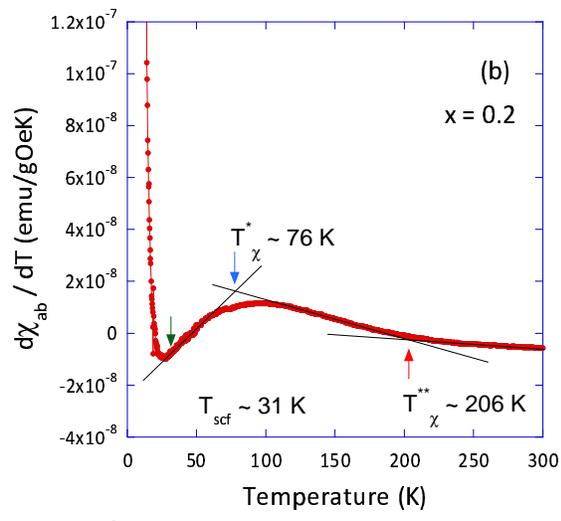
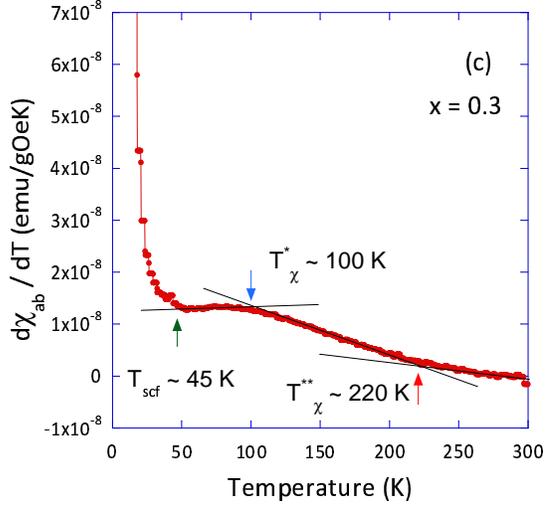
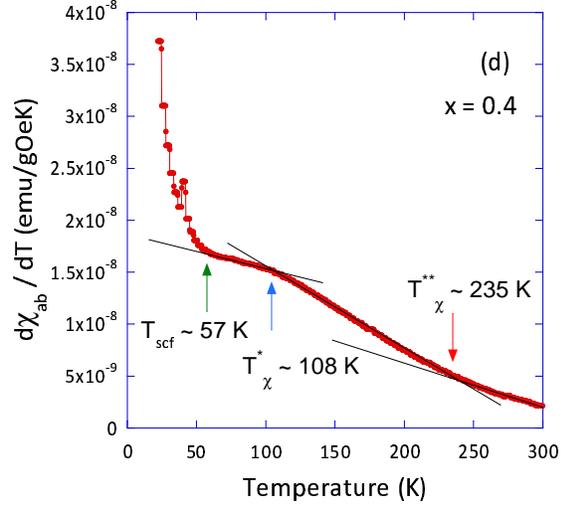
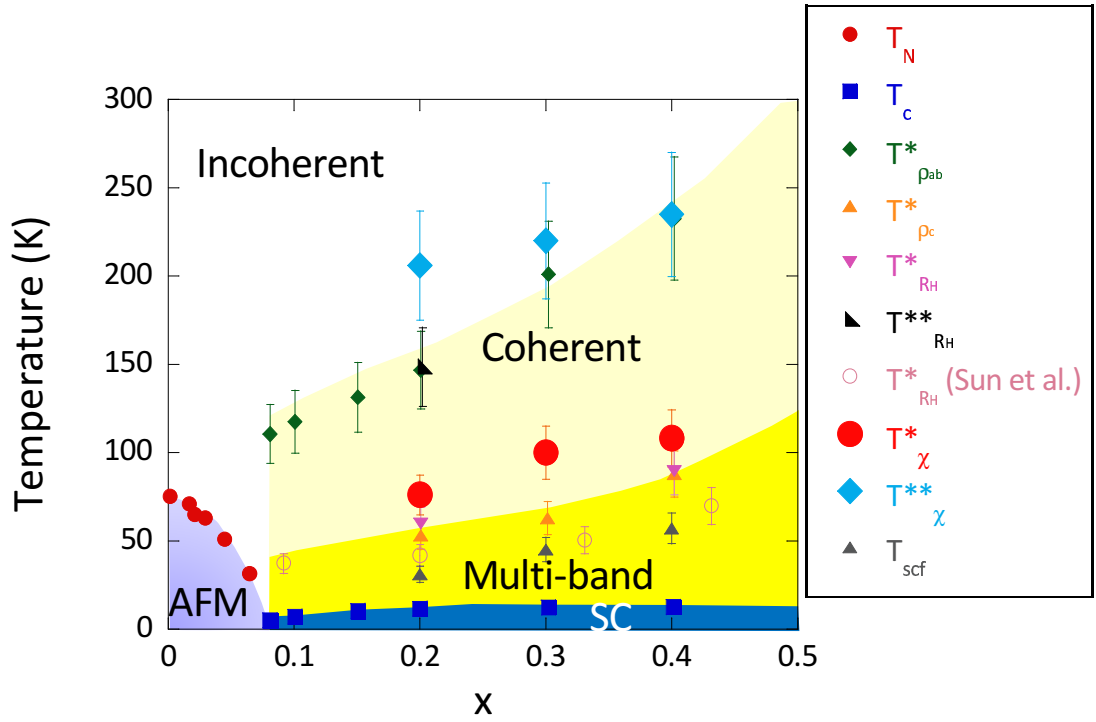
\end{document}